\documentclass{aa}
\usepackage{graphics}
\usepackage{times}

\begin{document}

\thesaurus{08(08.09.2 \object{RX~J1826.2$-$1450}; 08.09.2
\object{LS~5039}; 08.09.2 \object{NVSS J182614$-$145054}; 13.25.5;
08.22.3; 13.18.5)}

\title{X-ray and radio observations of RX~J1826.2$-$1450/LS~5039}

\author{M. Rib\'o\inst{1}
\and P. Reig\inst{2,3}
\and J. Mart\'{\i}\inst{4}
\and J.M. Paredes\inst{1}
}

\institute{Departament d'Astronomia i Meteorologia, Universitat de 
Barcelona, Av. Diagonal 647, E-08028 Barcelona, Spain
\and Physics Department, University of Crete, P.O. Box 2208, 710 03, 
Heraklion, Greece
\and Foundation for Research and Technology-Hellas, 711 10, Heraklion,
Crete, Greece
\and Departamento de F\'{\i}sica, Escuela Polit\'ecnica Superior, 
Universidad de Ja\'en, Calle Virgen de la Cabeza, 2, E-23071 Ja\'en, Spain
}

\offprints{M. Rib\'o, mribo@mizar.am.ub.es}

\date{Received / Accepted}

\maketitle

\begin{abstract}

RX~J1826.2$-$1450/LS~5039 has been recently proposed to be a radio
emitting high mass X-ray binary. In this paper, we present an analysis of
its X-ray timing and spectroscopic properties using different instruments
on board the RXTE satellite. The timing analysis indicates the absence of
pulsed or periodic emission on time scales of 0.02--2000 s and 2--200 d,
respectively. The source spectrum is well represented by a power-law
model, plus a Gaussian component describing a strong iron line at 6.6 keV.
Significant emission is seen up to 30 keV, and no exponential cut-off at
high energy is required. We also study the radio properties of the system
according to the GBI-NASA Monitoring Program. RX~J1826.2$-$1450/LS~5039
continues to display moderate radio variability with a clearly non-thermal
spectral index. No strong radio outbursts have been detected after several
months.


\end{abstract}

\section{Introduction} \label{intro}

The star LS~5039 is the most likely optical counterpart to the X-ray
source RX~J1826.2$-$1450. Such an association was originally proposed by
Motch et al. (1997), hereafter M97, as a result of a systematic
cross-correlation between the ROSAT All Sky Survey (Voges et al. 1996) and
several OB star catalogues in the SIMBAD database. The unabsorbed X-ray
luminosity, at an estimated distance of 3.1 kpc, amounts to
$L_\mathrm{X}$(0.1--2.4~keV)~$\sim$~8.1$\times$10$^{33}$~erg~s$^{-1}$, and
the hardness of the source is well consistent with a neutron star or a
black hole, accreting directly from the companion's wind (M97). In the
optical, LS~5039 appears as a bright $V\sim$~11.2 star with an O7 V((f))
spectral type. Based on this evidence, M97 proposed the system to be a
high mass X-ray binary (HMXRB). 

In addition, this system has been found to be active at radio wavelengths.
Its radio counterpart (NVSS~J182614$-$145054) is a bright, compact and
moderately variable radio source in excellent sub-arcsecond agreement with
the optical star (Mart\'{\i} et al. 1998). All these facts point to the
peculiar nature of RX~J1826.2$-$1450/LS~5039, and suggest a classification
among the selected group of radio emitting HMXRB. 

In order to explore how this source behaves compared to other members of
its class (e.g. \object{Cygnus~X-1}, \object{LS~I+61$^{\circ}$303} and
\object{SS~433}), we have analyzed the corresponding X-ray data from the
All Sky Monitor (ASM) and the Proportional Counter Array (PCA) on board
the satellite Rossi X-ray Timing Explorer (RXTE). In Sect.~\ref{timing} we
present an X-ray timing analysis based on both the ASM and the PCA
instruments. The ASM data are suitable to study the long-term (days to
months) temporal behavior of the source, whereas X-ray variability on
shorter time scales (seconds to hours) is better investigated with the
PCA. In Sect.~\ref{spectral} a PCA spectroscopic analysis is presented,
with the different spectral models that fit the data being examined and
discussed. 

In the radio domain, RX~J1826.2$-$1450/LS~5039 was included at our request
in the list of radio sources routinely monitored at the Green Bank
Interferometer (GBI)\footnote{The Green Bank Interferometer is a facility
of the USA National Science Foundation operated by NRAO in support of the
NASA High Energy Astrophysics programs.}. At the time of writing, the
radio light curves cover $\sim$~4 months of observations. In
Sect.~\ref{radio} we present the GBI radio data so far acquired with some
discussion on the source variability and spectral index properties. 
Finally, we conclude in Sect.~\ref{comparison} with a brief comparative
discussion of RX~J1826.2$-$1450/LS~5039 versus other radio loud HMXRB. 

Hereafter, we will refer to the source as RX~J1826.2$-$1450 when
discussing the X-rays. In the optical/radio context the LS~5039
designation will be preferred.

\begin{figure*}[htb]
\mbox{}
\vspace{8.7cm}
\includegraphics{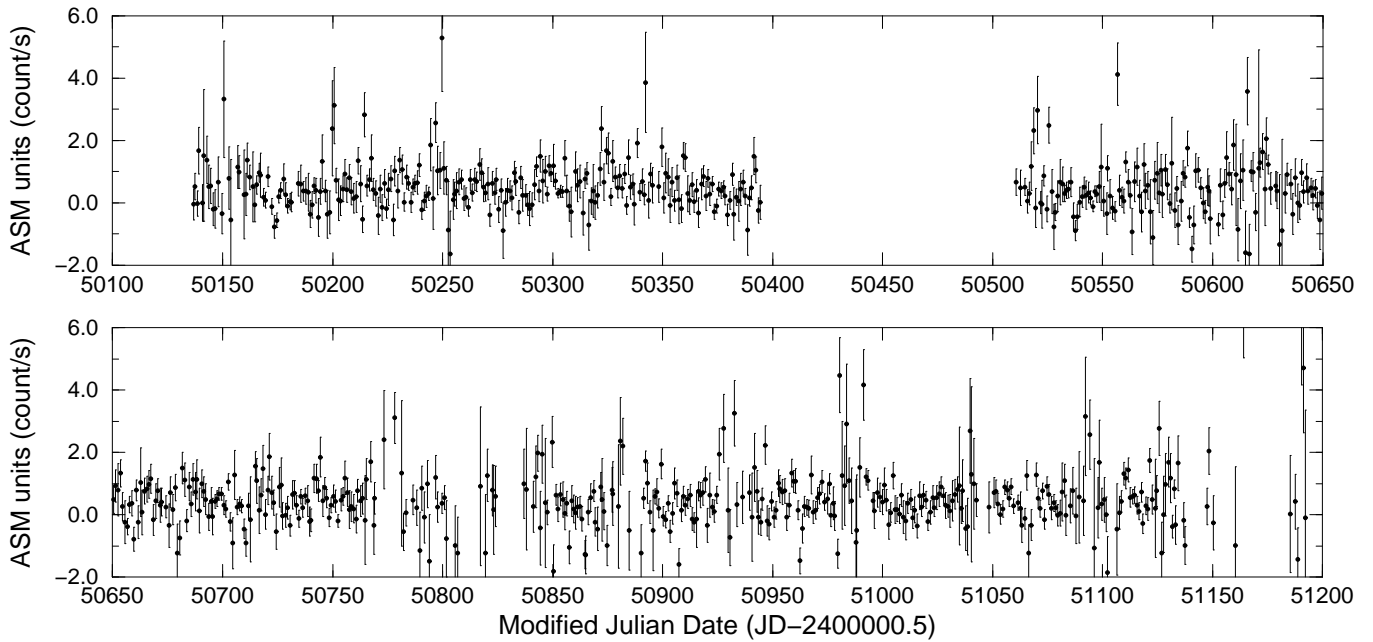}
\caption[]{ASM one-day average light curve of 
RX~J1826.2$-$1450 in the 1.5--12 keV energy band.}
\label{asm_lc}
\end{figure*}

\section{X-ray timing analysis} \label{timing}

\subsection{The ASM/RXTE data} \label{asm}

The ASM database analyzed in this paper spans for more than two and a half
years (1996 February--1998 November) and contains nearly 800 daily flux
measurements in the energy range 1.5--12 keV. Each data point represents
the one-day average of the fitted source fluxes from a number (typically
5-10) of individual ASM dwells, of $\sim$ 90 s each (see Levine et al.
1996 for more details). The one-day average light curve is shown in
Fig.~\ref{asm_lc}. The big gap between Modified Julian Date (MJD)
$\sim$~50400 and $\sim$~50500 corresponds to the passage of the Sun close
to the source during the first year of observations. This gap repeats the
following two years (near MJD 50800 and 51150), but it happens to be less
severe and a few flux measurements were then possible. 

Most of time, the source is at the threshold of ASM detectability.
Nevertheless, we have searched for possible periodicities in the range
from 2 to 200 d. The methods employed were the Phase Dispersion
Minimization (PDM) (Stellingwerf 1978) and the CLEAN algorithm (Roberts et
al. 1987). Our approach here is essentially the same as in Paredes et al.
(1997) when analyzing the periodic behavior in the X-ray light curve of
LS~I+61$^{\circ}$303. 

After applying both the PDM and CLEAN methods to the ASM data, a period of
$\sim$~52.7 d stands prominently. This periodicity corresponds to the
detection of some kind of active events that appear rather evident at
first glance in Fig.~\ref{asm_lc}. Nevertheless, a careful inspection of
the data reveals a suspicious detail. All those active events take place
when the data by dwell coverage is rather poor (less than 5 dwells per
day), thus reducing the statistical significance of the corresponding
one-day average. For some instrumental reason, the ASM coverage becomes
poorer than normal every $\sim$~53 d or so and, in the case of a weak
X-ray source like RX~J1826.2$-$1450, this can affect somehow the period
analysis. Therefore, the $\sim$~52.7 d period is very likely to be an
instrumental artifact. Indeed, after removing all daily points resulting
from less than 5 dwells ($\sim$~20\% of total), the timing analysis
reveals no significant period in the range from 2 to 200 d. 

\subsection{The PCA/RXTE data} \label{pca}

Additional observations were made with the PCA instrument on 1998 February
8 and 16. The total on-source integration time was 20 ks. The PCA is
sensitive to X-rays in the energy range 2--60 keV and comprises five
identical co-aligned gas-filled proportional counter units (PCUs),
providing a total collecting area of $\sim$~6500 cm$^2$, an energy
resolution of $<$ 18 \% at 6 keV and a maximum time resolution of 1$\mu$s.
Our analysis was carried out in the interval 3--30 keV since the PCU
windows prevent the detection of photons below $\sim$~2.5 keV, whereas
above 30 keV the spectrum becomes background dominated. 

\begin{figure}[htb]
\mbox{}
\vspace{10.3cm}
\includegraphics{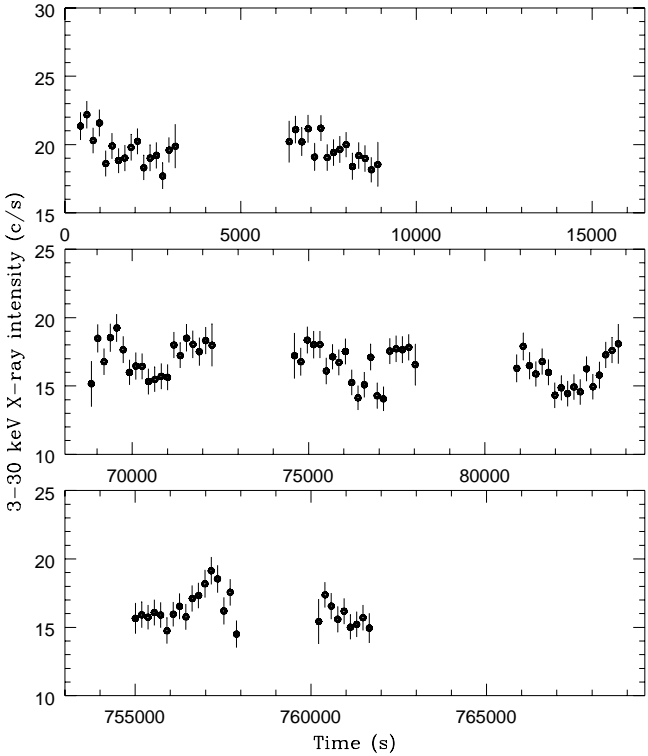}
\caption[]{3--30 keV light curve of RX~J1826.2$-$1450 covering the entire
PCA observation. Time 0 is JD 2\,450\,853.047 and the bin size is 180 s}
\label{pca_lc}
\end{figure}

Good time intervals were defined by removing data taken at low Earth
elevation angle ($<$ 8$^{\circ}$) and during times of high particle
background. An offset of only 0.02$^{\circ}$ between the source position
and the pointing of the satellite was allowed, to ensure that any possible
short stretch of slew data at the beginning and/or end of the observation
was removed. Table~\ref{tabx} shows the journal of the PCA observations,
while the light curve of the entire observation is presented in
Fig.~\ref{pca_lc}. 

\begin{table}
\begin{center}
\caption[]{\label{tabx} Log of the PCA observation of RX~J1826.2$-$1450}
\begin{tabular}{ccccc}
\hline
Date     & Start time & Stop time & Luminosity$^a$ \\
         & (TT)       & (TT)      & (erg s$^{-1}$) \\
\hline 
08/02/98 & 01:05:33   & 04:18:14  & 6.4 $\times$ 10$^{34}$ \\
08/02/98 & 20:07:34   & 00:57:14  & 5.3 $\times$ 10$^{34}$ \\
16/02/98 & 18:42:25   & 20:35:14  & 5.3 $\times$ 10$^{34}$ \\
\hline
\end{tabular}
\end{center}
$^a$ in the energy range 3--30 keV and for an assumed distance of 3.1 kpc\\
\end{table}

\begin{figure}[htb]
\mbox{}
\vspace{5.9cm}
\includegraphics{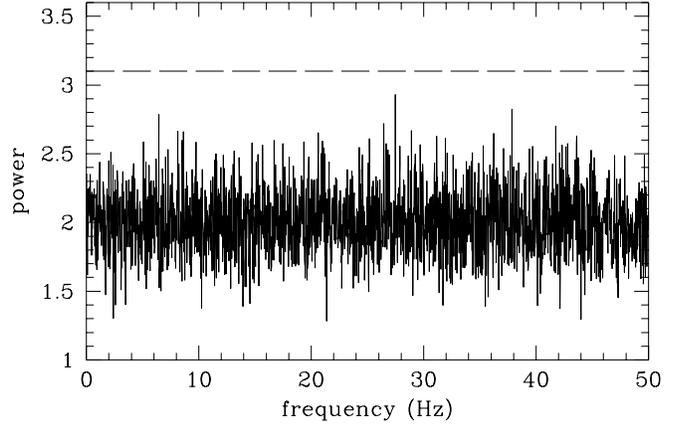}
\caption[]{Characteristic power spectrum of RX~J1826.2$-$1450. The dashed
line represents the 95\% confidence detection limit}
\label{pds}
\end{figure}

The PCA observations were used to study the time variability on various
time scales. Continuous stretches of clean data were selected from the
light curve of the entire observation. To reduce the variance of the noise
powers, these intervals were divided up into segments of 8192 bins each,
with a bin size of 10 ms. Then the power density spectra for each segment
were calculated and the results averaged together. Fig.~\ref{pds} shows
the characteristic power spectrum in the frequency range 0.01--50 Hz. The
dashed line represents the 95\% confidence detection limit (van der Klis
1989). As it can be seen no power exceeds this value; the distribution of
powers is flat at a level of 2, consistent with Poissonian counting
statistics. Following van der Klis (1989), we can set a 95\% upper limit
of 60\% on the {\em rms} of a pulsed source signal in the range 0.01--50
Hz. This relatively high limit is a consequence of the faintness of the
source. On longer time scales, longer intervals ($\sim$ 3200 s) were
considered but no evidence for pulsations was found either. 

Likewise, we folded the light curve onto a set of trial periods with the
FTOOLS software package (a technique very similar to the PDM) and looked
for a peak in the $\chi^2$ {\em versus} period diagram. None of the peaks
found were statistically significant enough. Thus, we conclude that no
coherent periodicities were detected in the range $\sim$~0.02 to
$\sim$~2000 s. 

The mean X-ray intensity in the energy range 3--30 keV shows a slight
decreasing trend with 19.7$\pm$0.2 count s$^{-1}$ at the beginning of the
observation (upper panel of Fig.~\ref{pca_lc}) compared to 16.6$\pm$0.1
count s$^{-1}$ and 16.2$\pm$0.2 count s$^{-1}$ for the middle and bottom
panels of Fig.~\ref{pca_lc}, respectively. The fractional {\em rms} of the
3--30 keV light curve corresponding to the entire observation is 9\%.

The fact that no X-ray pulsations have been found in RX~J1826.2$-$1450 is
consistent with the proposed idea that radio emission and X-ray pulsations
from X-ray binaries seem to be statistically anti-correlated (Fender et
al. 1997), i.e., no X-ray pulsar has ever shown significant radio
emission.

\section{X-ray spectral analysis} \label{spectral}
 
\subsection{Spectral fitting} \label{fitting}

\begin{figure}[htb]
\mbox{}
\vspace{11.0 cm}
\includegraphics{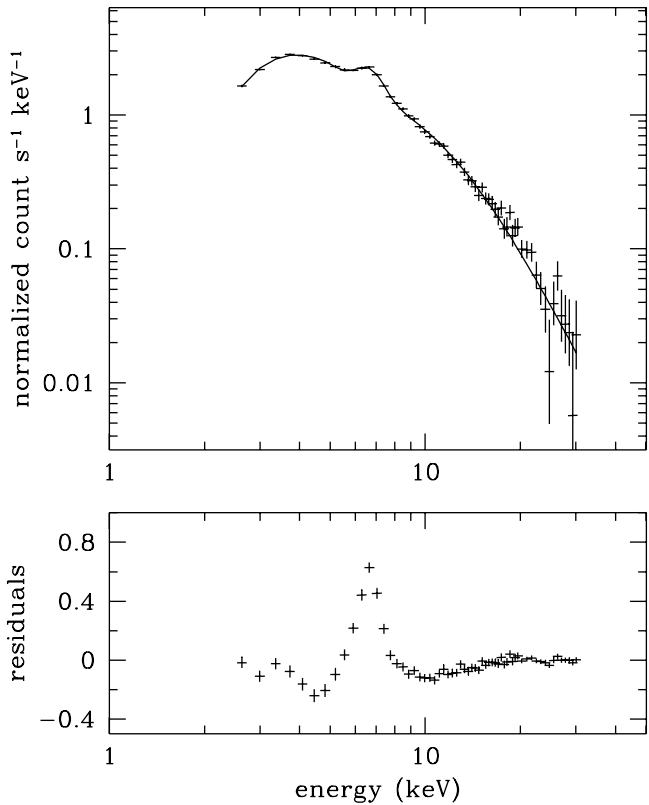}
\caption[]{PCA spectrum of RX~J1826.2$-$1450. The continuous line
represents the best-fit power-law model plus a Gaussian component for the
iron line. When this component is omitted the line shows up clearly in the
residuals}
\label{pca_spec}
\end{figure}

Since the light curve of the entire observation does not show sharp
features, i.e., there is no significant spectral change throughout the
observation, we obtained one average PCA energy spectrum from the complete
observation (Fig.~\ref{pca_spec}). 

Acceptable fits of the X-ray continuum were obtained with an unabsorbed
power-law model, giving a reduced $\chi^2$=1.14 for 56 degrees of freedom
(dof). A multi-colour disk model, as expected from an optically thick
accretion disk (Mitsuda et al. 1984) plus a power-law gave a reduced
$\chi^2_{\nu}$=1.11 for 53 dof. The best-fit results are given in
Table~\ref{fit}. Bremsstrahlung and two blackbody component models did not
fit the data. Although the addition of a blackbody component to the
power-law formally produces an acceptable fit, the value of the blackbody
normalization was very low, with the error bar close to zero. In fact, an
F-test shows that the inclusion of a blackbody component is not
significant. 

\begin{table}
\begin{center}
\caption[]{\label{fit} Spectral fit results for the power-law model plus 
a Gaussian component for the iron line. Uncertainties are given 90\% 
confidence for one parameter of interest. The spectrum was fitted in the
energy range 3-30 keV}
\begin{tabular}{lc}
\hline
Parameters                         & Fitted values \\
\hline
$N_\mathrm{H}$ ($\times 10^{21}$ cm$^{-2}$) & 2$^{+1}_{-2}$ \\
$\alpha$                           & 1.95$\pm$0.02 \\
$E_\mathrm{line}$(Fe)~(keV)        & 6.62$\pm$0.04 \\
$EW_\mathrm{line}$(Fe)~(keV)       & 0.75$\pm$0.06 \\
$FWHM_\mathrm{line}$(Fe)~(keV)     & 0.9$\pm$0.2   \\
$\sigma_\mathrm{line}$(Fe)~(keV)   & 0.39$\pm$0.08 \\
$\chi^2_r$ (dof)                   & 1.14 (56)     \\
\hline
\end{tabular}
\end{center}
\end{table}

The most salient feature that appears in the spectrum of RX~J1826.2$-$1450
is a strong iron line at $\sim$~6.6 keV (Fig.~\ref{pca_spec}). A Gaussian
fit to this feature gives a line centered at 6.62$\pm$0.04 keV, with an
equivalent width ($EW$) of 0.75$\pm$0.06 keV and a $FWHM$ of 0.9$\pm$0.2
keV. The high $EW$(Fe) value indicates that a large amount of
circumstellar matter is present in the system. Unfortunately, the PCA
energy resolution prevents from distinguishing between a broad line or two
narrow components. 

A hydrogen column density of
$\sim$~(2$^{+1}_{-2}$)~$\times$~10$^{21}$~cm$^{-2}$ is found from the fit. 
This value is, however, not very well constrained. In fact, it is
consistent with zero. The difficulty in constraining the hydrogen column
density from our X-ray data can be attributed to the fact that the
interstellar gas mainly absorbs X-ray photons with energies lower than
2--3 keV, i.e., outside of the energy range considered here. Nevertheless,
it agrees with the value obtained from optical observations. From the
$\lambda$~4430 and $\lambda$~6284 interstellar bands M97 found
$E(B-V)$~=~0.8$\pm$0.2. Using the relation
$N_\mathrm{H}$~=~5.3~$\times$~10$^{21}$~cm$^{-2}~E(B-V)$ (Predehl \&
Schmitt 1995), we obtain that
$N_\mathrm{H}\sim$~(4$\pm$1)~$\times$~10$^{21}$~cm$^{-2}$, which is
consistent, within the errors, with the X-ray observation value. 

\subsection{A black hole or a neutron star?} \label{nature}

As mentioned above, no pulsations are found in the X-ray flux of
RX~J1826.2$-$1450/LS~5039. At the time of writing this paper approximately
55\% of the optically identified massive X-ray binary systems are pulsars,
rising to 67\% if suspected HMXRB are included. About 75\% of the
identified and suspected X-ray pulsars have spin periods below 100
seconds. The detection of pulsations would rule out a black hole
companion.

Interestingly, unlike typical X-ray pulsars, the energy spectrum of
RX~J1826.2$-$1450 shows no exponential cut-off at high energy despite that
significant emission is seen up to 30 keV. The absence of both, pulsations
and high energy cut-off may indicate that the compact companion is a black
hole rather than a neutron star. The persistent non-thermal radio emission
of RX~J1826.2$-$1450/LS~5039 is also reminiscent, among others, of the
classical black hole candidate (BHC) Cygnus~X-1. However, the data are not
conclusive as counter-examples of these properties can be found. For
example, the system LS~I+61$^{\circ}$303, which seems to contain a neutron
star and does not exhibit pulsations or cut-off in the 10--30 keV range.
The multi-colour disk model, although formally fitting the data, does not
help either. First, given the low luminosity
($L_\mathrm{X}<$10$^{35}$~erg~s$^{-1}$ in the energy range 2--10 keV) the
system would be in the so-called low-state. We would not expect then to
detect a strong soft component. Second, the fit provides an unrealistic
value of the disk internal radius of $R_\mathrm{in}\cos^{1/2}(\theta) \sim
0.3$ km. Finally, the lack of detected pulsations may be just due to the
faintness of the X-ray emission in view of the rather high upper limit
found in Sect.~\ref{pca} for pulsed emission. 

Unfortunately, the source is too faint at energies above 30 keV to be
detected with the HEXTE instrument. Thus, we cannot confirm from the
present data whether the hard tail that characterizes the energy spectrum
of BHCs at high energies is indeed present. In any case, the issue of a
possible black hole in RX~J1826.2$-$1450/LS~5039 is likely to be set in
the future by obtaining a spectroscopic mass function in the optical.

\section{Radio observations} \label{radio}

\subsection{Radio variability properties of LS~5039} \label{radiovar}

\begin{figure*}[htb]
\mbox{}
\vspace{7.9cm}
\includegraphics{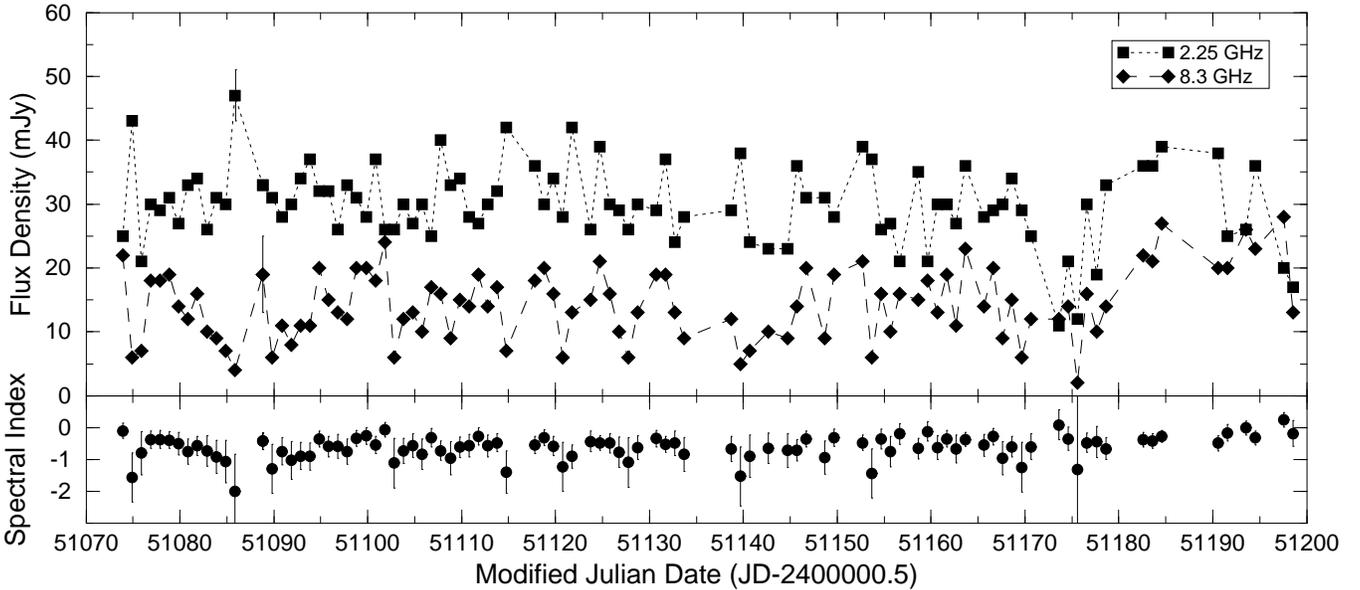}
\caption[]{{\bf Top:} GBI radio light curves of LS~5039 at the frequencies
of 2.25 and 8.3 GHz. Representative $\pm$~1$\sigma$ error bars have been
plotted. {\bf Bottom:} The corresponding spectral index. Error bars are
also $\pm$~1$\sigma$. The source behaves as moderately variable and has a
clear non-thermal spectral index}
\label{gbi_lc}
\end{figure*}

The radio observations reported here consist of daily flux density
measurements with the GBI within the GBI-NASA Monitoring Program. The
source was observed at the frequencies of 2.25 and 8.3 GHz. In
Fig.~\ref{gbi_lc}, we show the corresponding flux density light curves
during the first $\sim$~4 months of monitoring so far available. The
bottom panel displays the spectral index $\alpha$ (where
$S_{\nu}\propto\nu^\alpha$) computed between these two frequencies. 

The source was found to be detectable at radio wavelengths throughout all
the time. This behavior is better evident at 2.25 GHz where the source is
brighter. The average GBI flux densities and their respective {\it rms}
are $S_{\rm{2.25~GHz}}=31(\pm5)$ mJy and $S_{\rm{8.3~GHz}}=14(\pm5)$ mJy.
The typical day-to-day variability in the GBI data does not exceed
$\sim30$~\%. There may be however some exceptions, such as for example
around Modified Julian Date (MJD) 51075, 51086 and 51176. Here, the flux
density of LS~5039 seems to have varied by more than a factor of $\sim$~2
on less than one day. Excluding these episodes, the source never exhibited
well defined radio outburst events. In general terms, the observed radio
behavior confirms the early suggestions by Mart\'{\i} et al. (1998)
concerning the persistent and moderately variable nature of the radio
emission. 

Timing analysis of radio light curves has proven to be in some cases a
useful tool to detect orbital periods. For example, the orbital period
signatures of 26.5 and 5.6 d are visible in LS~I+61$^{\circ}$303 and
Cygnus~X-1, respectively (Taylor \& Gregory 1984; Pooley et al. 1998). We
have thus searched for long-term periodicities in the GBI data of LS~5039.
Given the span of the radio observations, the search was restricted
between 2 and 50 d. The methods used were again the PDM and CLEAN,
mentioned in Sect.~\ref{asm}. Unfortunately, no convincing period was
detected in this process. A longer time span is likely to be required
before a reliable search can be attempted for this relatively weak radio
source (specially at 8.3 GHz).

\subsection{Non-thermal radio spectrum and brightness temperature}
\label{radiospec}

From the GBI data, the weighted average spectral index is found to be
$\alpha=-0.5^{+0.2}_{-0.3}$. This value is also in good agreement with the
results by Mart\'{\i} et al. (1998) obtained a few months before, thus
suggesting that the non-thermal radio spectrum is a persistent property of
the source. 

In addition to negative spectral indices, the brightness temperature
estimates for LS~5039 clearly yield to non-thermal values, hence
supporting the mechanism of synchrotron radiation for that source. The
apparent one-day variability observed in the GBI data around MJD 51075,
51086 and 51176 would imply, from light time travel arguments, that the
emitting region is smaller than about $2.6\times10^{15}$ cm. If we assume
a 3.1 kpc distance to the source, the corresponding angular size is found
to be $\theta\leq$~0${\rlap.}^{\prime \prime}$05, yielding a lower limit
of $T_b\geq6\times10^6$ K (at 2.25 GHz). This lower limit is not far from
the $T_b\geq4\times10^6$ K estimate by Mart\'{\i} et al. (1998), based on
the unresolved nature of the source with the VLA.

\begin{table*}
\begin{center}
\caption[]{\label{hmxrb} Average properties of radio emitting HMXRBs from 
ASM/RXTE and GBI data} 
\begin{tabular}{lccccccc}
\hline
HMXRB & $L_{\mathrm{X}}$(1.5--12~keV) & $L_{\mathrm{rad}}$(0.1--100 GHz) &
$\alpha$ & Compact object & Companion & Orbital    & Distance \\
           &(erg~s$^{-1}$)            & (erg~s$^{-1}$)             &
($S_{\nu}\propto\nu^\alpha$) & &      & period (d) & (kpc) \\
\hline

LS~5039    & $\sim5\times10^{34}$~~~~~~ & $1.0\times10^{31}$~~~~~~ &
$-0.5$~~~~~~ & ?              & O7~V((f))      & ?    & 3.1$^{a}$ \\

Cygnus~X-1 & $\sim8\times10^{36}$ (*)   & $1.1\times10^{31}$~~~~~~ &
$0.1$~~~     & black hole     & O9.7~Iab       & 5.6  & 2.5$^{b}$ \\

LS~I+61$^{\circ}$303 & $\sim4\times10^{34}$~~~~~~ & $0.9\times10^{31}$ (*)&
$-0.3$ (*)   & neutron star ? & B0Ve           & 26.5 & 2.0$^{c}$ \\

SS~433     & $\sim7\times10^{35}$ (*)   & $3.2\times10^{32}$ (*)   &
$-0.7$ (*)   & neutron star ? & OB ?           & 13.1 & 4.8$^{d}$ \\ 

\hline
\end{tabular}
\end{center}
(*) Only data during quiescence has been considered\\
$^{a}$ Motch et al. 1997\\
$^{b}$ Penninx 1989\\
$^{c}$ Frail \& Hjellming 1991\\
$^{d}$ Vermeulen et al. 1993\\
\end{table*}

\section{Comparison to other radio loud HMXRB} \label{comparison}

SS~433, LS~I+61$^{\circ}$303 and the BHC Cygnus~X-1 are classical examples
of HMXRB with detectable radio emission (Penninx 1989), and all of them
are also under GBI monitoring. In order to facilitate the comparison of
LS~5039 to these sources, we have summarized some relevant parameters in
Table~\ref{hmxrb}. They include: the quiescent X-ray luminosity, as
derived from ASM/RXTE data (extrapolated from PCA/RXTE in the case of
LS~5039 because the signal to noise ratio is much higher); the weighted
average radio luminosity and spectral index in quiescence, both based on
GBI data; the nature of the compact object; the spectral type of the
companion; the orbital period of the binary system and the distance to the
source. 

As it can be seen in Table~\ref{hmxrb}, the radio luminosity of LS~5039 is
very similar to that of Cygnus~X-1. This is, of course, provided that the
distance adopted is correct. Their respective spectral indices seem to be
intrinsically different, both sources being persistent at radio
wavelengths. Cygnus~X-1 also experiences strong X-ray variability due to
changes in its state, that we are not aware of in our source. The
LS~I+61$^{\circ}$303 radio properties during quiescence are also very
comparable to those of LS~5039, as well as their respective X-ray
luminosities. In contrast, LS~I+61$^{\circ}$303 undergoes radio outbursts
every $\sim$~26.5~d, the orbital period of the system, while LS~5039 never
had strong outbursts during the GBI observations. The X-ray and radio
luminosities of SS~433 in quiescence are much higher than those of
LS~5039. However, we notice that their $L_{\mathrm{rad}}/L_{\mathrm{X}}$
ratios are practically the same. 

In general terms, the average properties of LS~5039 do not deviate
extraordinarily from those of other radio loud HMXRB. Since even the well
accepted members of this class are not an homogeneous group, the belonging
of LS~5039 to this category appears as very plausible to us.

\section{Conclusions} \label{conclusions}

We have presented a general overview of the X-ray and radio emission
properties of the massive X-ray binary RX~J1826.2$-$1450/LS~5039. Our
X-ray and radio results are mostly based on long term (few months)
monitorings of the source, with our main conclusions being:

\begin{enumerate}

\item In the X-rays, a timing analysis has been performed showing neither
pulsed nor periodic emission on time scales of 0.02--2000 s and 2--200 d,
respectively. The X-ray spectrum has been found to be significantly hard
(up to 30 keV), with no cut-off required. It can be fitted satisfactorily
with a power-law plus a strong Gaussian iron line.

\item At radio wavelengths, the GBI monitoring confirms the long-term
persistence of the RX~J1826.2$-$1450/LS~5039 radio emission in time scales
of months, always with a non-thermal synchrotron spectrum. The day-to-day
variability continues to be moderate most of the time ($\la30$~\%), and no
strong radio outbursts have been observed.

\item The classification of RX~J1826.2$-$1450/LS~5039 among the radio loud
HMXRB group is reinforced. Although some specific differences with other
members of this class do exist, noticeable similarities can be found.

\end{enumerate}

\begin{acknowledgements}
We thank Ron Remillard for useful discussion about the ASM data. We also
thank Iossif Papadakis for his help in the timing analysis of the PCA
data. This paper is partially based on quick-look results provided by the
ASM/RXTE team and data obtained through the HEASARC Online Service of
NASA/GSFC.
We acknowledge detailed and useful comments from an anonymous referee.
M.R. is supported by a fellowship from CIRIT (Generalitat de Catalunya,
ref. 1999~FI~00199).
P.R. acknowledges support via the European Union Training and Mobility of
Researchers Network Grant ERBFMRX/CP98/0195.
J.M. is partially supported by Junta de Andaluc\'{\i}a (Spain).
J.M.P and J.M. acknowledge partial support by DGICYT (PB97-0903).

\end{acknowledgements}

\end{document}